\documentstyle[aps,epsfig,multicol]{revtex} 

\newcommand{\be}{\begin{equation}}
\newcommand{\ee}{\end{equation}}
\newcommand{\bea}{\begin{eqnarray}}
\newcommand{\eea}{\end{eqnarray}}

\newcommand{\lp}{\left(}
\newcommand{\rp}{\right)}

\begin{document}
\draft

\title{
Dynamical spin-electric coupling in a quantum dot}
\author{L.\,S. Levitov,$^{1}$ E.\,I. Rashba$^{1,2}$\cite{Rashba*}}
\address{$^{1}$\ Department of Physics, 
 Center for Materials Sciences \& Engineering,
Massachusetts Institute of Technology, Cambridge, MA 02139;\\
$^{2}$\  Department of Physics, SUNY at Buffalo, Buffalo, NY\,14260 }
\maketitle
\begin{abstract} 
  Due to the spin-orbital coupling in a semiconductor quantum dot, 
a freely precessing 
electron spin produces a time-dependent charge density. This creates
a sizeable electric field outside the dot, leading to promising applications
in spintronics. 
The spin-electric coupling can be employed 
for non-invasive single spin detection
by electrical methods.
We also consider a spin relaxation mechanism due to long-range
coupling to electrons in gates and elsewhere in the system,
and find a contribution comparable to, and in some cases 
dominant over previously discussed mechanisms.
\end{abstract}
\begin{multicols}{2}

Because of the spin-orbit (SO) interaction, a precessing electron spin 
in a semiconductor produces a time-dependent oscillating electric field 
along with a magnetic field. This effect is weak in
single molecules, 
because the SO coupling is small in the
inverse Dirac gap $2m_{\rm e}c^2\approx 1\,{\rm MeV}$.
In semiconductors it is enhanced,
since the SO splitting of the upper valence band
($0.3\,{\rm eV}$ in GaAs, $0.9\,{\rm eV}$ in InSb)
can reach or even exceed the energy gap size. The SO effects
are further reinforced by low symmetry\cite{RS91},
allowing for a strong coupling of the electron spin to static 
and time-dependent electric fields. 

We propose to employ
the electric field produced by a freely precessing 
electron spin in a quantum dot for non-invasive
single spin 
detection, which may have promising applications in spintronics\cite{spintr}. 
A different idea, based on charge transport through the dot,
was put forward by Engel and Loss\,\cite{EngelLoss}. 
The spin-electric coupling considered below leads to novel physical effects,
including a new mechanism 
of spin relaxation complementary to that discussed recently
by Khaetskii and Nazarov\,\cite{KN}.

Low symmetry is crucial for this. 
While in spherically symmetric systems, such as atoms, 
the SO-induced orbital magnetization currents do not produce 
electric dipole or higher multipole moments, 
in a less symmetric system  
the orbital currents can generate 
a time-varying electric field accompanying 
spin precession. 
To clarify the underlying physics and 
to simplify the calculations, 
we consider a 
quantum dot with highly anisotropic confinement,
such as a dot created within a quantum wire. 


The SO interaction arises in this case 
from two separate contributions: 
the confinement-enhanced bulk inversion asymmetry (BIA, Dresselhaus) 
and the structure inversion asymmetry (SIA, Rashba)\,\cite{asym,dot}. 
The resulting SO interaction can be written as 
${\cal H}_{\rm SO}=\alpha\sigma_2\hat k$, 
where $\hat k=-i\hbar\partial_x$ 
and $\sigma_2$ is a suitably chosen Pauli matrix. 
This form of SO interaction has been used for dislocations\cite{disl} 
and quantum wires\,\cite{wire}. The total $1D$
Hamiltonian is 
%
\be
{\cal H}_0 = \hbar^2\hat k^2/2m^\ast+U(x)+\alpha\sigma_2\hat k
\label{eq1}
\ee
with $U(x)$ the confining potential. Without loss of generality we choose 
external magnetic field ${\bf B}\parallel \hat {\bf z}$, so that
the Zeeman interaction is
${\cal H}_Z = -\mu B\sigma_3$. (There is no orbital coupling to
vector potential in $D=1$.)

We eliminate ${\cal H}_{\rm SO}$ from ${\cal H}_0$ 
by a canonical transformation with a unitary matrix $S=e^{i\sigma_2x/2\xi}$.
Here the length $\xi=\hbar^2/2m^\ast\alpha$ coincides with 
the characteristic size of the Datta and Das device\,\cite{DD}.
The transformation shifts $\hat k$ by $\sigma_2/2\xi$ and moves
the SO coupling to the Zeeman term:
\bea
{\cal H}_0&=&\hbar^2\hat k^2/2m^\ast+U(x)-m^\ast\alpha^2/2\hbar^2,\nonumber\\
{\cal H}_Z&=&-\mu B[\sigma_3\cos(x/\xi)-\sigma_1\sin(x/\xi)].
\label{eq2}
\eea
For a weak magnetic field,
the Zeeman term can be treated 
as a perturbation.
In a symmetric potential, $U(x)=U(-x)$, the 
mean value of the second term in ${\cal H}_Z$ vanishes 
and the spin Hamiltonian 
projected on two Kramers-conjugate states becomes diagonal: 
\be
{\cal H}_n =\langle n|{\cal H}_Z|n\rangle=-\mu B\langle n|\cos(x/\xi)|n\rangle\,\sigma_3,
\label{eq3}
\ee
where $n$ labels orbital wave functions, $\psi_n(x)$. The Zeeman splitting
in Eq.\,(\ref{eq3}) depends on the SO coupling 
$\alpha$. 

For narrow gap ${\rm A_3B_5}$ quantum wells, the typical values of $\alpha$ 
originating from SIA are about $10^{-9}\,{\rm eV\, cm}$\,\cite{alp1}. 
However, larger values of $\alpha$ up to $3\cdot10^{-9}\,{\rm eV\, cm}$ 
for ${\rm In_{0.75}Ga_{0.25}As/In_{0.75}Al_{0.25}As}$ 
heterojunctions\cite{alp2} and $6\cdot10^{-9}\,{\rm eV\, cm}$ 
for ${\rm In_{0.52}Al_{0.48}As/In_xGa_{1-x}As}$ structures\cite{alp3} 
were reported more recently. 
Both experiment\cite{Grundler} and theory\cite{calc} 
indicate that the interface asymmetry makes an important, 
and maybe even dominant, 
contribution to $\alpha$, that can be varied by system design. 
With $\alpha\approx6\cdot10^{-9}\,{\rm eV\, cm}$ and $m^\ast/m\approx 0.05$, 
we estimate the characteristic length scale as $\xi\approx 13\,{\rm nm}$. 
The dependence of ${\cal H}_n$ on the SO coupling can thus be significant 
for quantum dots of size comparable or larger than $\xi$.
(In diffusive dots the SO effects are controlled by the
ratio of level spacing to 
the SO scattering spin flip rate\,\cite{HalperinMarcus}.)

Now we consider the time-dependent electric charge density arising due 
to electron spin precession. It is given by the off-diagonal 
in spin element of the density matrix
$N^{(n)}_{\downarrow\uparrow}(x,t)= 
e^{-i\omega_{Z,n}t}N^{(n)}_{\downarrow\uparrow}(x)$
with the $n$'th orbital state Zeeman frequency $\omega_{Z,n}$ 
defined by Eq.~(\ref{eq3}). Here
\be
N^{(n)}_{\downarrow\uparrow}(x)=
\langle {\bar\Psi}_{n\downarrow}(x)|\Psi_{n\uparrow}(x)\rangle_{\rm spin}
\label{eq4}
\ee
with
$\Psi_{n\sigma}(x)$ the exact Zeeman-split Kramers 
doublet spinor wave functions.
The partial trace in Eq.\,(\ref{eq4}) is taken over spin. 
To the first order in $\omega_Z$ one obtains
\be
N^{(n)}_{\downarrow\uparrow}(x) = 2\mu B\sum_{n^{\prime}\neq n}
{{\psi_n(x)\psi_{n^{\prime}}(x)}\over{E_n-E_{n^{\prime}}}}
\langle {n^{\prime}}|\sin(x'/\xi)|n\rangle,
\label{eq5}
\ee
and the corresponding dipole moment equals
\be
P_n=-2e\mu B \sum_{n^\prime\neq n}{{\langle n|x'|n^\prime\rangle
\langle n^\prime|\sin(x''/\xi)|n\rangle}
\over{E_n-E_{n^\prime}}}.
\label{eq6}
\ee
To complete this general discussion, let us consider free spin precession
in a quantum dot holding $2n+1$ electrons, one electron with unpaired 
spin at the $n$'th level, with all lower states $E_{n^{\prime}}<E_n$
fully filled. For the average spin of the $n$'th level 
at an angle $\theta$ to the magnetic field, the time-dependent
dipole is
\be
{\bf P}(t)= \sin\theta \, P_n \, \cos(\omega_{Z,n} t +\varphi)\,\hat{\bf x}
\label{eq:P(t)}\ee
with $\omega_{Z,n}$ and $P_n$ given by Eq.\,(\ref{eq3}) and Eq.\,(\ref{eq6}).

Now we focus on two practically interesting confinement models:
a square well with hard walls, describing a quantum wire segment, and 
a parabolic quantum dot. For a wire segment of length $L$,
using sinusoidal standing wave states,
the Zeeman frequency (\ref{eq3}) is
\be
\hbar\omega_{Z,n}(\gamma)=
2\mu B\,\frac{\pi^2n^2\sin\gamma}{(\pi^2n^2-\gamma^2)\gamma},\quad \gamma=L/2\xi
\label{eq7}
\ee
with $n\geq 1$. (Zeros in the denomenator do not cause divergence because of 
$\sin\gamma$.) The electric dipole (\ref{eq6}) is
\bea\label{eq8}
&&P_n(\gamma)=-eL\,(\mu B/\Delta_n) \\
&&\times \sum_{n^\prime-n={\rm odd}}
{(2^8/\pi^4)(2n+1)\,(nn^\prime)^2\gamma\cos\gamma\over(n^2-{n^\prime}^2)^3
[(n^2\!+\!{n^\prime}^2\!-\!(2\gamma/\pi)^2)^2\!-\!4(n{n^\prime})^2]}
\nonumber
\eea
with $\Delta_n=E_{n+1}-E_n=(2n+1)\pi^2\hbar^2/2m^\ast L^2$ the separation between 
the energy levels $E_{n+1}$ and $E_n$. 
The sum, evaluated exactly for $n=1$ and $\gamma\ll 1$, gives
\be
P_1=
{{15-\pi^2}\over 8\pi^2}\,{{\mu B}\over \Delta_1}\,{L\over\xi}\,eL
\label{eq9}
\ee
This result is similar to the matrix element 
of electric-dipole transitions in 3D donor centers\,\cite{RS64}. 
The two factors multiplying the geometric dipole $eL$ 
in (\ref{eq9})
have the following meaning.
The factor $\mu B/\Delta_1\ll 1$ reflects that the matrix element 
of $x$ between two Kramers-conjugate states vanishes at $B=0$
due to the time-reversal symmetry. 
The factor $L/\xi\ll 1$ makes $P_1$ vanish
at zero SO coupling. 
Despite the small factors, the electric dipole 
can still be much larger than the Bohr's magneton
$\mu=\frac12 e\lambdabar_C$, where $\lambdabar_C$ is 
the electron Compton length. 

For a parabolic dot with a confining potential $U(x)=m\omega^2x^2/2$, 
the Zeeman frequency (\ref{eq3}) is 
\be
\hbar\omega_{Z,n}(\gamma_2)=2\mu B e^{-\gamma_2/2}L_n(\gamma_2), \quad
\gamma_2={\textstyle\frac12}(x_0/\xi)^2
\label{eq10}
\ee
where $n\geq 0$, $x_0=(\hbar/m\omega)^{1/2}$, 
and $L_n$ are Laguerre polynomials. Similar to Eq.~(\ref{eq7}),
the frequency (\ref{eq10}) is a  
sign-changing function 
of the SO coupling $\alpha$, vanishing at $\alpha\rightarrow\infty$. 
Summation in Eq.~(\ref{eq6}), performed exactly 
using harmonic oscillator selection rules, gives
\be
P_n(\gamma_2)= e x_0 (\mu B/\hbar\omega)\,(2\gamma_2)^{1/2} e^{-\gamma_2/2}
L_n(\gamma_2).
\label{eq11}
\ee
For the ground state, the dipole moment (\ref{eq11}) is similar to 
$P_1$ of Eq.~(\ref{eq9}) with $x_0$ replacing $L$.

\begin{figure}
\centerline{\psfig{file=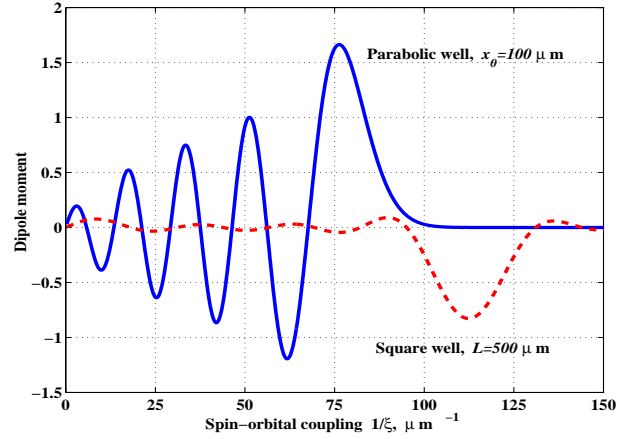,width=3.2in,height=2.3in}}
\vspace{0.1cm}
        \caption[]{
Electric dipole of a Zeeman-split state {\it vs} SO coupling.
The state  $|n\rangle$ with $n=9$ is used.
The dipole (\ref{eq8}) for a square well
is scaled by $\frac13 eL\,(\mu B/\Delta_5)$; the dipole 
(\ref{eq11}) for a parabolic well is scaled by $ex_0\,(\mu B/\hbar\omega)$.
        }
\label{fig:P(xi)}
\end{figure}

The dipole moments (\ref{eq8}) and (\ref{eq11}) for 
the two models (Fig.~\ref{fig:P(xi)}) behave as follows.
For a square well $P_n(\gamma)$ has a peak at $\gamma\approx\pi n$
of width $\delta\gamma\simeq 2\pi$ independent on $n$. The peak is
positive or negative depeding on the parity of $n$,
with weak oscillations on both sides. For a parabolic well
the oscillations of $P_n(\xi)$ increase in amplitude as $\xi^{-1}$ 
up to $\gamma_2\approx 4n$
and then abruptly disappear as $P_n(\gamma_2)$ drops.

\begin{figure}
\centerline{\psfig{file=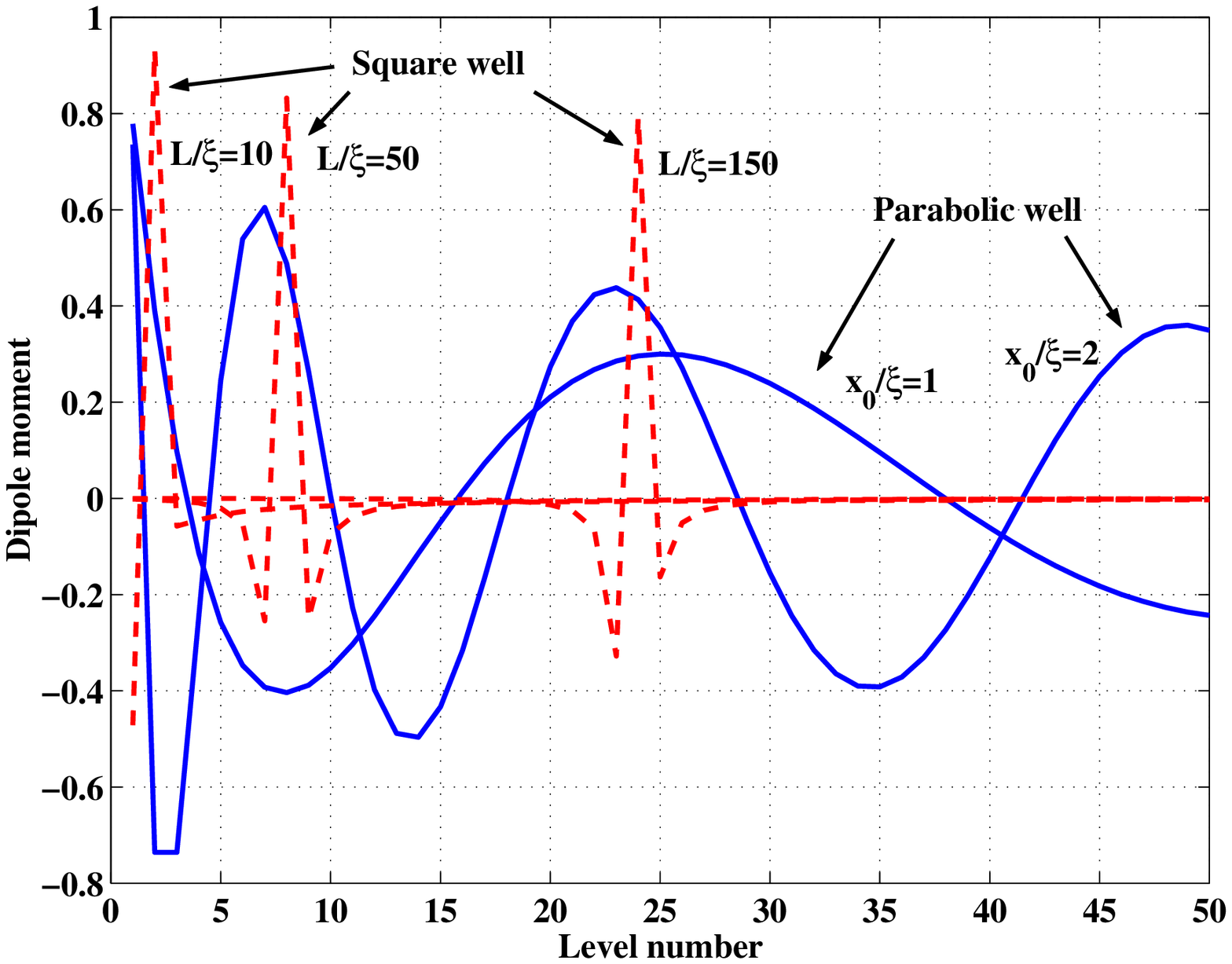,width=3.2in,height=2.3in}}
\vspace{0.1cm}
        \caption[]{
Electric dipole scaled as in Fig.\,\ref{fig:P(xi)} {\it vs} level number.
        }
\label{fig:P(n)}
\end{figure}

The dependence of the dipole on the level number is quite interesting 
(Fig.\,\ref{fig:P(n)}). For the square well, since
the dipole (\ref{eq8}) peaks near $\gamma=\pi n$, 
$P_n$ is large only for specific levels, while for 
the harmonic potential it is a smooth oscillatory fucntion 
of level number. This dramatic difference is explained as follows. 
In a square well each wavefunction is characterized by a specific wavelength, 
equal to $2L/n$ for the $n$'th level, while in a harmonic potential the 
wavelength is position-dependent. The SO effect will be strong 
when the wavelength matches the spatial period $4\pi\xi$ 
of the matrix $e^{i\sigma_2x/2\xi}$ used to gauge out the SO interaction.
Thus one expects the dipole in a square well to be large for the states with
$2L/n=4\pi\xi$, which is exactly the above condition $\gamma=\pi n$.
In a harmonic potential, on the other hand, there should be no specific
levels with enhanced dipole.

A much stronger spin-electric coupling arises
for non-Kramers states brought to degeneracy 
at the Zeeman energy
$\mu B$ matching level separation  
$E_n-E_{n^\prime}$. For SO-split avoided crossings of levels 
with opposite spin and different orbital wavefunctions,
the SO-induced electric field is not small
in the factor $\mu B/\Delta$ appearing in Eq.~(\ref{eq9}) 
due to time-reversal symmetry at $B=0$. 
The SO-split level crossings in small 
elongated dots were reported by Rokhinson {\it et al.}\cite{Rokh} 
and in 2D dots by Folk {\it et al.}\cite{Folk}

For such a pair of states $|\psi_n\!\!\uparrow\rangle$, 
$|\psi_{n'}\!\!\downarrow\rangle$ the energy separation in the absence 
of the SO coupling is
$\Delta=E_n-E_{n^\prime}-2\mu B$. The avoided crossing of levels split by the 
SO matrix element $V$ is described by $\delta E=(\Delta^2+4V^2)^{1/2}$.
The off-diagonal charge density matrix element is
\be
N_{+-}^{nn^\prime}(x)=u_{+}u_- \psi^2_n(x)+v_{+}v_- \psi^2_{n^\prime}(x),
\label{eq12}
\ee
where
$
u_\pm=\left[(\delta E\pm\Delta)/2\delta E\right]^{1/2},~~
v_\pm=\pm u_\mp
$
are the components of 
the two states participating in the avoided crossing. 
(The quantities $\psi_n$, $\psi_{n'}$ and $V$ 
are real due to the absence of magnetic orbital coupling
in $D=1$.) Evaluating 
$u_+u_-=-v_+v_-= V/(\Delta^2+4V^2)^{1/2}$
we obtain
\be
N_{+-}^{nn^\prime}(x)=
\frac{V}{(\Delta^2+4V^2)^{1/2}}\,
(\psi^2_n(x)-\psi^2_{n^\prime}(x))
\label{eq12a}
\ee
This charge density 
oscillates with the frequency $\delta E/\hbar$.

We note that the distribution (\ref{eq12a})
possesses a dipole moment only for an asymmetric confining potential,
$U(x)\neq U(-x)$, while the quadrupole moment 
$Q_{nn^\prime} = e\int x^2N_{\downarrow\uparrow}^{nn^\prime}(x)\,dx$
exists even for symmetric 
dots:
\be
Q_{nn^\prime}
={eV\over (\Delta^2+4V^2)^{1/2}}
\int x^2[\psi_n^2(x)-\psi_{n^\prime}^2(x)]dx.
\label{eq14}
\ee
Near the resonance, 
$\Delta \approx V$, the quadrupole moment $Q_{nn^\prime}$ {\it contains 
no small factors} and is controlled by the integrand. 
For a parabolic confinement potential
\be
Q_{nn^\prime}=ex_0^2\,(n-n^\prime)\,V/(\Delta^2+4V^2)^{1/2}
\label{eq15}
\ee
The factor $ex_0^2n$ has the scale of the quadrupole moment 
of the $n$'th quantum state. 
The enhancement of the electrical signal from a spin precessing 
between two non-Kramers levels resembles 
a similar effect for the electric dipole spin resonance 
at acceptor centers\,\cite{RS64}. 

{\it Electrical detection of a single electron spin precession}
is attractive because of high sensitivity of
electrical measurements\cite{electrometry}. 
Moreover, electrical detection 
can be performed locally, 
e.g. by a single-electron transistor (SET)\cite{localSET} or just by 
measuring the time-dependent potential induced on the gates
around the dot holding spin. 

To estimate the magnitude of the effect, we consider an electron
in a square well defined in a quantum wire. The potential at a distance
$r$ from the dot has the order of magnitude 
$\varphi(r)\approx P_1/\epsilon r^2$
with $P_1$ given by Eq.~(\ref{eq9}). For an estimate, taking 
$L=20\,{\rm nm}$, $r=50\,{\rm nm}$, $\epsilon=13$, and 
$L/\xi\approx1$, 
we obtain
\be
\varphi(r)\approx 
(\mu B/\Delta_1)\cdot 0.1\,{\rm mV}
\label{eq:phi(r)}
\ee
The detection of a signal with the amplitude (\ref{eq:phi(r)}) 
oscillating at the Zeeman frequency is certainly feasible\,\cite{electrometry}.

We expect a stronger effect in
the two other situations considered above.
In a parabolic dot\cite{shallow} the dipole is typically larger 
than in a square well, mainly because of its smooth dependence 
on level number (Fig.\,\ref{fig:P(n)}). 
In the case of non-Kramers level crossing, the effect is enhanced 
due to the absence of the small factor 
$\mu B/\Delta_n$. The leading effect is dipolar for asymmetric 
and quadrupolar for symmetric dots. 
Although exact estimates are problematic because of 
a large number of independent parameters, we expect
the effect in this case to be stronger 
than (\ref{eq:phi(r)}). 
Electric signal arising near avoided crossings can also 
be used to
detect level intersections at constant charge in the Coulomb 
blockade regime.

If spin precession is excited by a resonant external field (ESR),
care should be taken to separate the spin-electric signal
from the excitation. One possibility is to employ a pulsed 
ESR excitation and detect precession 
signal ``ringing'' after each pulse. 
Slow spin relaxation times of up to few microseconds reported by Fujisawa 
{\it et al.} \cite{Fujisawa} (see also
\cite{spintr}) will simplify the task.

Another possible way is to use 
{\it thermal excitation} of spin precession, 
i.e. to work at relatively weak 
magnetic fields, $1/\tau_2\ll\mu B\alt k_{\rm B}T$, with $\tau_2$ the spin 
dephasing time. Since both spin states, as well as
their superpositions, are populated in thermal equilibrium, 
no external ESR excitation is required in this case.
The spin-electric signal will give rise to a narrow band noise forming
a peak of width $1/\tau_2$ at the Zeeman frequency. 
The noise peak value is $\approx \varphi(r)\tau_2^{1/2}$.
With $\tau_2=1\,{\rm \mu s}$ we estimate the peak noise signal
as $(\mu B/\Delta)\cdot 0.1\,{\rm \mu V/Hz}^{1/2}$.

A narrow band noise of this form, with a peak at $\omega=\omega_Z$, 
was discovered experimentally by Manassen {\it et al.}\cite{Mana}
in STM current detected near paramagnetic centers on Si surfaces. 
One can speculate, based on the above, that the SO-induced electric field
modulates the tunnel barrier for STM current in vacuum,
effectively turning STM in a spin detector.
A recent explanation 
by Balatsky and Martin\cite{Bal}, also based on SO coupling, predicts current modulation 
$\delta I/I\propto I$, while our mechanism remains effective 
even for $I\rightarrow 0$. Therefore, these mechanisms can be easily 
distinguished experimentally.

The spin-electric coupling discussed above leads to
{\it a new mechanism of spin relaxation} in a quantum dot surrounded 
by metallic electrodes.
The low frequency electric field of the dipole ${\bf P}(t)$ 
penetrates inside the metal, where it can transfer 
the excitation energy $\hbar\omega_Z$ to Fermi system.
This mechanism is {\it dissipationless} because quasiparticles 
acquire energy during passage near the surface and then 
dissipate it somewhere far away. 

The interaction 
takes place within the screening length $r_s$
near metal surface, where the screened potential is
$\tilde\varphi(z,\rho)=-\partial_{z}\varphi(\rho) r_s e^{-z/r_s}$
(here $\rho$ and $z$ are the coordinates along and 
perpendicular to the surface, and
$\partial_{z}\varphi(\rho)$ is normal derivative).
The spin relaxation rate can be found from {\it the Golden Rule}:
\be
W=\frac{e^2}{\hbar}\int\!\!\int
2\,{\rm Im}\,{\cal K}(\omega_Z,{\bf r}_1,{\bf r}_2) 
\tilde\varphi({\bf r}_1)\tilde\varphi({\bf r}_2)
d^3{\bf r}_1
d^3{\bf r}_2
\label{eq:GoldenRule}
\ee
with ${\cal K}(\omega_Z,{\bf r}_1,{\bf r}_2)$
the two-point density correlator in the metal and ${\bf r}\equiv (z,\rho)$.
Finite temperature adds 
the factor 
$(1-e^{-\hbar\omega_Z/k_BT})^{-1}$. 
Below we consider $k_{\rm B}T\ll \hbar\omega_Z$.

In the case of specular boundary conditions 
on metal surface, using the method of images,  
the correlator ${\cal K}$ near the surface
can be related with that in the bulk:
${\cal K}_{12}=K_{12}+K_{12'}$,
where $2'$ is a mirror image of the point 2.
We then rewrite Eq.\,(\ref{eq:GoldenRule}) as
$
W=\frac{e^2}{\hbar}\sum_{\bf k}|\tilde\varphi({\bf k})|^2
\,{\rm Im}\, K(\omega_Z,{\bf k})
$. 
In a clean metal, using
${\rm Im}\, K(\omega,{\bf k})=
\frac{\pi}2\nu\omega/|{\bf k}|v_F$, with $\nu$ the density of states,
%
\bea
&&W=\frac{e^2\nu}{4\hbar}\omega_Z
\int (\partial_{z}\varphi)^2_{z=0} d^2\rho
\int \lp\frac{2r_s^2}{1+k^2r_s^2}\rp^2\frac{ dk}{v_F|k|}
\nonumber\\
&&\approx (e^2/h v_F)\ln(r/r_s)\,(Pr_s/er^2)^2\,\omega_Z
\label{eq:Wclean}
\eea
The log arises due to particles incident at small angles 
that interact stronger 
with the screened dipole field.


For diffusive metal, with
$K(\omega,{\bf k})\!=\!-\nu D{\bf k}^2/(D{\bf k}^2\!-\!i\omega)$,
\bea
&&W=\frac{e^2\nu r_s^4}{2\pi^2\hbar D}
\int\,{\rm Im}\,\frac{i\omega_Z|\partial_{z}\varphi({\bf q})|^2}{({\bf q}^2-i\omega_Z/D)^{1/2}}\,d^2{\bf q}\\
&&\simeq 
\cases{(P r_s/r^2)^2(\omega_Z/D)^{1/2}/\hbar, & $\omega_Z\gg D/r^2$\cr
(P r_s/r^2)^2\,(\omega_Zr\!/\!D\hbar)\ln(D\!/\!r^2\omega_Z), & $\omega_Z\ll D/r^2$}
\eea
We find that diffusion enhances $W$ by the number $N$ of returns 
to the surface during the coherence time, estimated as
$N\simeq \,{\rm min}\,[(\omega_Z\tau)^{-1/2}, r/l]$,
with $\tau$ and $l$ the elastic mean free time and path. 
Here the first bound
ensures that the electric field is nearly constant in time,
while the second bound restricts the displacement of a diffusing
quasiparticle to the distance $r$ from the dot.

Estimating the relaxation rate (\ref{eq:Wclean}) with the above 
parameter values, one has $W\simeq (\mu B/\Delta)^2\cdot 10^{-6}\omega_Z$. 
Recently, 
electron spin relaxation due to  coupling
to fluctuating magnetic fields\,\cite{KN} 
and to nuclear spins\,\cite{nuclei}
was considered. 
Although a direct numerical comparison is difficult due to
a wide spread of parameter values, Eq.(19) of Ref.\,\cite{KN} 
gives a number similar to ours obtained
for $\mu B/\Delta\simeq 10^{-2}$. In the case of non-Kramers 
level crossing, due to the absence of the factor $\mu B/\Delta$,
the electric mechanism can dominate.
Generally, the electrical and magnetic spin relaxation rates 
depend on different combinations of parameters and, 
therefore, should be considered as complementary mechanisms. 

In summary, the SO-induced electric field around  
a freely
precessing spin can be employed for single spin detection. It may also
significantly contribute to spin relaxation. The reverse
spin-electric effect is also of interest
in view of spintronics applications. It can serve as a mechanism for
independent spin monitoring and control in 
different dots by 
local electric field sources.

This work is supported by the MRSEC Program
of the National Science Foundation under Grant No. DMR 98-08941
(LL)
and by DARPA/SPINS, the Office of Naval Research Grant No. 000140010819 (ER).

\end{multicols} 

\begin{references}
\bibitem{shallow} 
Similar enhancement should occur in any shallow dot.
\end{references}
\end{document}